\documentclass[twocolumn,showpacs,showkeys,preprintnumbers,amsmath,amssymb]{revtex4}
\usepackage{subfigure}
\usepackage{graphicx}% Include figure files
\usepackage{dcolumn}% Align table columns on decimal point
\usepackage{bm}% bold math

\begin{document}
\newcommand{\be}{\begin{equation}}
\newcommand{\ee}{\end{equation}}
\newcommand{\mean}[1]{\left\langle #1 \right\rangle}
\newcommand{\abs}[1]{\left| #1 \right|}
\newcommand{\set}[1]{\left\{ #1 \right\}}
\newcommand{\la}{\langle}
\newcommand{\ra}{\rangle}
\newcommand{\lb}{\left(}
\newcommand{\rb}{\right)}
\newcommand{\norm}[1]{\left\|#1\right\|}
\newcommand{\RA}{\rightarrow}
\newcommand{\tet}{\vartheta}
\newcommand{\eps}{\varepsilon}

\preprint{APS/123-QED}

\title{Anti-deterministic behavior of discrete systems \\ that are
  less predictable than noise}

\author{Krzysztof Urbanowicz\footnote{Corresponding author:\\
        Krzysztof Urbanowicz\\Max Planck
Institute for the Physics of Complex Systems\\
 N\"{o}thnitzer Str. 38\\ D--01187 Dresden, Germany\\
 Tel.: +49-351-871-2217\\
 Fax: +49-351-871-1999\\
 E-mail:urbanow@mpipks-dresden.mpg.de}}
 \affiliation{Max Planck
Institute for the Physics of Complex Systems\\
 N\"{o}thnitzer Str. 38\\ D--01187 Dresden, Germany}
\author{Holger Kantz}
 \affiliation{Max Planck
Institute for the Physics of Complex Systems\\
 N\"{o}thnitzer Str. 38\\ D--01187 Dresden, Germany}
\author{Janusz A. Ho{\l}yst}%
\affiliation{Faculty of Physics and Center of Excellence for
Complex Systems Research,
\\Warsaw University of Technology\\ Koszykowa 75, PL--00-662 Warsaw,
Poland}
 \affiliation{Max Planck
Institute for the Physics of Complex Systems\\
 N\"{o}thnitzer Str. 38\\ D--01187 Dresden, Germany}

\date{\today}

 \begin{abstract}
\par
We present a new  type of deterministic dynamical behaviour that is
less predictable than white noise. We call it
anti-deterministic (AD) because  time series  corresponding to the
dynamics of such
systems do not generate deterministic lines in Recurrence Plots for
small thresholds. We
show that although the dynamics is chaotic in the sense of exponential
divergence of nearby initial conditions and although
some properties of AD data are similar to
white noise, the AD dynamics  is in fact less predictable than noise
and hence is different from pseudo-random number generators.
\end{abstract}
%Use showkeys class option if keyword
\maketitle

\section{Introduction}
 \par
Determinism in the strict sense is defined by the existence of a
unique solution of the initial value problem: Given an equation of
motion, then the motion generated by it is said to be deterministic if
the solution evolving from a given initial condition depends
exclusively on the latter, i.e., it is uniquely determined (for all
future times) by fixing the initial condition\cite{Schuster}. This property is
usually contrasted by stochastic motion, where random noises enter the
equation of motion, such that the initial condition alone is
insufficient to fix the future evolution. One additionally needs to know
the realization of the sequence of noise inputs. Without this, one can
only make probabilistic statements about the evolving solutions.

In the above restricted sense, the dynamics which we introduce
here is in fact deterministic. However, when speaking about
determinism, one often considers the aspect of recurrence:
Determinism implies that when, in the course of time, the system
returns to a state which it has assumed before, the evolution will
repeat itself precisely. If in addition (as it is most often the
case) the terms in the equation of motion depend smoothly on the
state vector, then the return to a {\sl similar} state will cause
a {\sl similar} future evolution, at least on short times (in
chaotic systems, this time is related to the inverse of the
Kolmogorov-Sinai entropy). This follows from a simple
linearization of the equations around a given reference state and
is the basis of many time series tools in reconstructed phase
spaces\cite{KantzSchreiber}. In particular, this is the basis of
the famous ``Lorenz method of analogues''\cite{Lorenz} or, more
technically speaking, the zeroth order prediction
scheme\cite{FarmerSidorowich}: If one wants to predict the short
time future of a deterministic system, the simplest method is to
look for situations in the past which are very similar to the
present one, and to assume that (because of the above discussed
properties of determinism) the future will be similar to what
followed the similar situation in the past. In many applications
to experimental data, such prediction schemes have in fact been
proven to work very successfully\cite{Abarbanel,KantzSchreiber}.
What we call anti-deterministic is exactly the opposite: whenever
our system comes to a state which is similar to some state of the
past, it will evolve in the most different way from the past, so
that past information is systematically misleading when trying to
make predictions.

This property would require a highly
non-smooth equation of motion. Instead, we will define the dynamics not
through an equation of motion but by a minimization problem. This is,
however, not too unusual, since one can convert every
differential equation initial value problem into a variational
problem, where the solution on a finite time interval is given by a
path which generates the extremum of some functional (compare the
principle of minimal action in classical mechanics\cite{Goldstein}).

A particular tool for the visual inspection of this aspect of
determinism is the recurrence
plot\cite{Eckmann}. A recurrence plot of a given time
series $\set{x_i},\;\; i=1,\ldots,N$ is the $N\times N$-matrix ${\bf R}$ with
the entries $R_{ij}=\Theta(\eps-|x_i-x_j|)$, where $\Theta$ is the
Heaviside step function. I.e., a matrix element is unity if the
correponding time series points are closer than $\eps$, and it is zero
else. Such a matrix can be easily represented graphically which is
called the recurrence plot, with the parameter $\eps$. Several
concepts for a quantitative evaluation of recurrence plots have been
proposed\cite{Zbilut,Casdagli,Potsdam}.

Based on the theorem of
Takens\cite{Takens}, determinism in the more general sense expresses
itself in time series data by the existence of line segments parallel
to the diagonal in this plot.
The lengths
of these lines for periodic or quasiperiodic systems are limited only
by the lengths $N$ of the corresponding time series
because there is no divergence
of nearby trajectories. The lines are much shorter for chaotic
motion because of a finite exponential divergence,
corresponding to the positive entropy of chaotic
systems\cite{Casdagli}. The lines themselves are representative of
unstable periodic orbits of the system.
Also for data stemming from uncorrelated stochastic processes (such as
white noise) there exists some small but nonzero probability
that a ``deterministic'' line can occur in a RP for a finite threshold
value $\eps$, since similar data segments may occur by chance.
The systems which we call anti-deterministic (AD) are designed to create
systematically less lines parallel to the diagonal  in
recurrence plots than white noise, and for certain recurrence
parameters they do not possess any line at all.

More specifically, we generate sequences of AD data by requiring
them not to create any line of length longer then $n$ in RP
maximizing the threshold $\eps$. We found several algorithms that
can produce such data in a deterministic way, and presumably one
can also introduce some randomness beyond the choice of random
initial conditions. In the next section we show one simple
procedure.  The AD motion does not belong to a class of chaotic
systems with a very high entropy (pseudo-random number
generators\cite{NumRec}). As it will become clear in the next
section, our construction of AD data requires an infinite memory
of the dynamical system, which corresponds to an infinite
dimensional phase space. There are neither (unstable) periodic
orbits nor is there any attractor. Data corresponding to AD
systems cover uniformly the whole admissible phase space.

\section{\label{sec.recipe}{A simple algorithm  for AD data generation}}
Working in discrete time, we generate the AD data iteratively. We
restrict the individual time series elements $x_i$ to be from the unit
interval, $x_i\in [0,1]\;\; \forall i$.
Let $\set{x_i}$ for $i=1,2,\ldots,N$ be a given sequence
representing the data up to time $N$.
The next time series element $x_{N+1}$ is the one that maximizes its
(suitably defined) distance to all previous points of the time series.
More precisely, $x_{N+1}$ maximizes the following utility function $U_{x}$:

\be
U_x=\min\limits_{i=d+m,d+m+1,\ldots,N}\prod_{\kappa=0}^{m}
 dis_d\lb i-\kappa,N+1-\kappa\rb\;.\label{eq.utilP}
\end{equation}
Here, $m$ and $d$ are
parameters of the algorithm, while $dis_d\lb i,N+1\rb$ is the Euclidean
distance between points $x_i$ and $x_{N+1}$ in a $d$-dimensional time
 delay embedding space:
\be dis_d(i,N+1)=\sqrt{\sum\limits_{k=0}^{d-1}
\norm{x_{i-k}-x_{N+1-k}}^2}\;.
\end{equation}
It can be useful to
impose periodic boundary conditions on the unit interval when
computing the distance, i.e., $\norm{x_i-x_j}:=
min(\abs{x_i-x_j},1-\abs{x_i-x_j})$.

For the simple choice $m=0$, Eq.(1) gives the distance (in the
$d$-dimensional time delay embedding space) between the ``new''
time series point and its closest neighbour among all past points,
which, by the choice of the new point, should be as large as
possible. For $m>0$, the antideterministic features which we study
below become more pronounced. In order to start the algorithm, an
initial sequence $\set{x_i}$ for $i=1,2,\ldots,n_0$ has to be
chosen.

\begin{figure}
\includegraphics[scale=0.35,angle=-90]{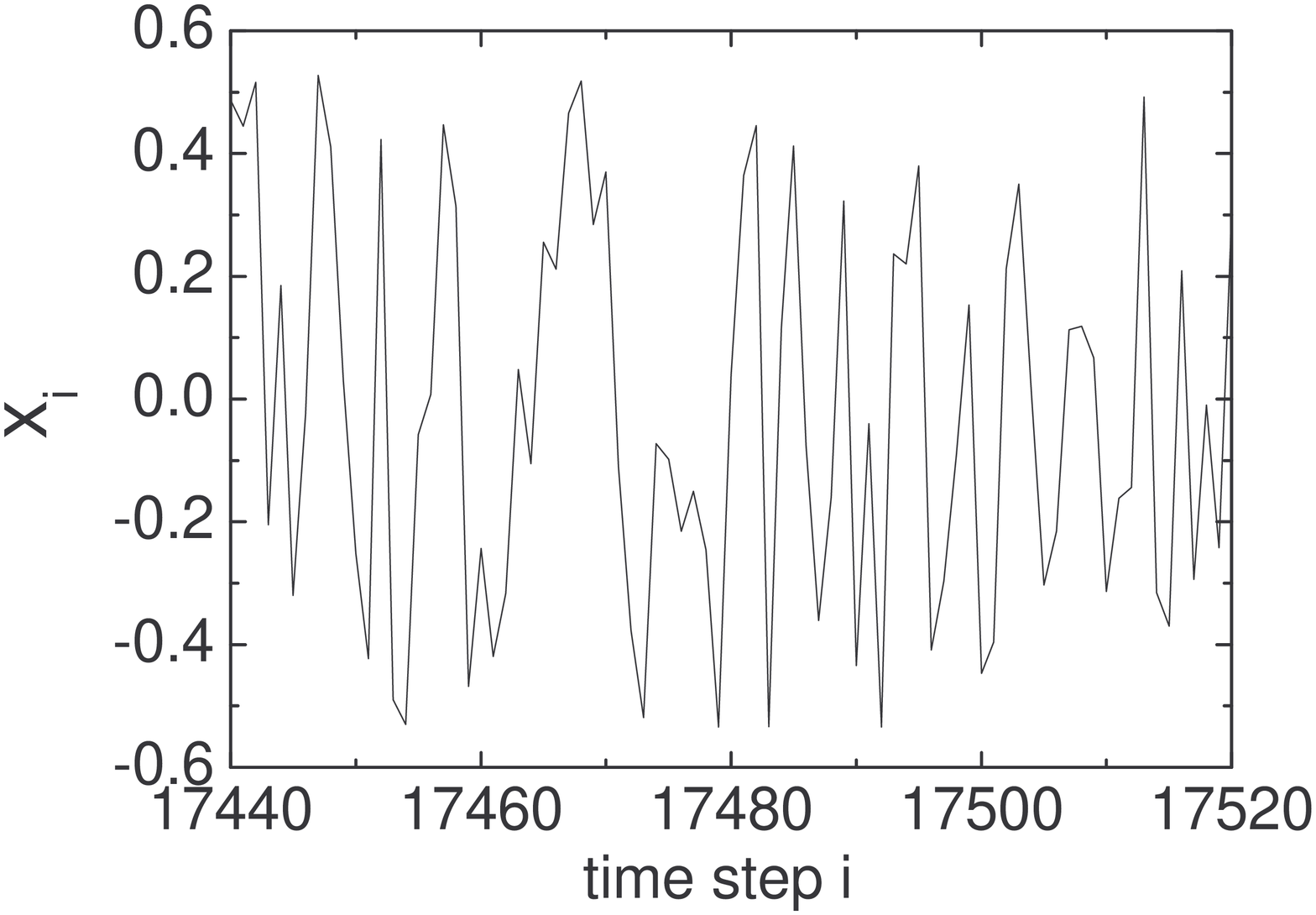}
\caption{\label{fig.anti} The example of AD  time series.}
\end{figure}

Our algorithm of  AD data generation is very
simple and one can imagine that this method can be modified in many
different ways, e.g., by modifying the function $U_x$. As long as the
main idea, namely that the new data point (in a time delay embedding sense of
Takens\cite{Takens}) is required to be far from all previous points,
is respected by $U_x$, the properties of the resulting data are all
very similar.
In Fig.~(\ref{fig.anti}) an example of an AD time series is shown.

\section{Properties}
\begin{figure}
\includegraphics[scale=0.35,angle=-90]{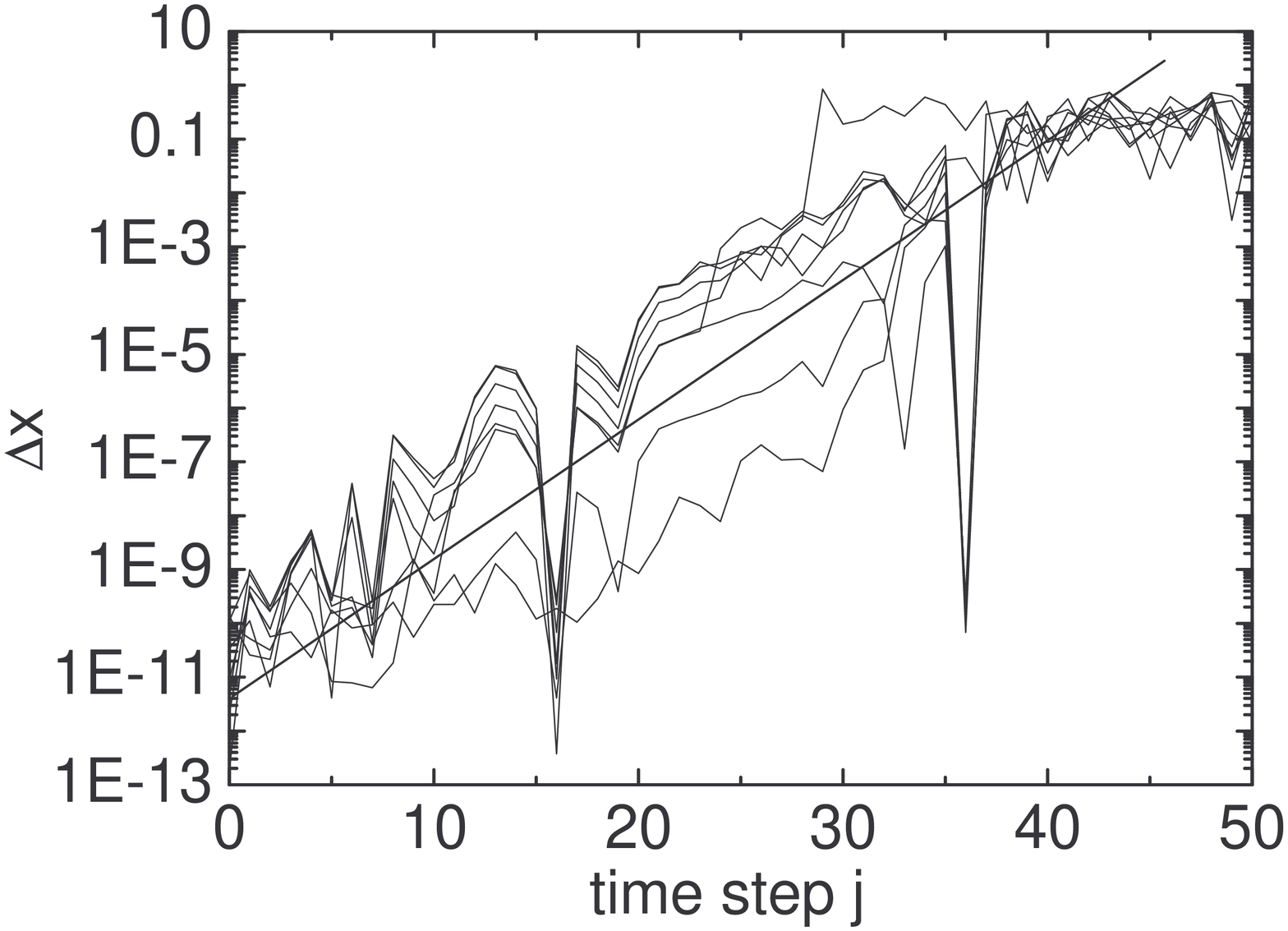}
\caption{\label{fig:divergence}The distance of time series elements
  $\Delta x=\norm{x_j-x'_j}$ evolving from nearby initial conditions ($n_0=50$,
  $m=4$, $d=2$, perturbation of the initial condition in the range of
  $10^{-10}$). The line guides the eye and corresponds to the behavior
  $e^{0.61 j}$.
}
\end{figure}

The requirement of anti-determinism translates into the fact that
there cannot be contracting and not even marginally stable
directions\cite{EckmannRuelle} in the dynamical system. This type
of behavior easily occurs in discrete time where  all directions
can be  divergent. In a conventional flow (continuous time system)
there is always one direction corresponding to a zero Lyapunov
exponent. Hence, if the generalization of AD to continuous time
dynamics is possible it will lead to a nowhere continuous signal
similar to white noise. Here, we restrict ourselves to the time
discrete version. An AD trajectory never forgets its initial
condition $\set{x_i}$, $i=1,2,\ldots,n_0$ and it is very sensitive
to their changes. As we can see from Fig.\ref{fig:divergence}, the
divergence of initially nearby solutions is, on average,
exponential, resembling low-dimensional chaos. This result is
consistent with our statement that AD data are usually no
high-entropic data such as those generated by linear congruential
random number generators. The specific realization also depends
sensitively on the parameters $m$ and $d$. The main feature of AD
data is the absence of RP lines of lengths larger than $d$ for
small values of $\eps$. In a conventional chaotic sequence,
despite the unpredictability in the long run, on short times
similar states evolve similarly. This is related to the existence
of unstable periodic orbits which are densely embedded in the
invariant set for relevant classes of chaotic
attractors\cite{Ott}. A close approach of the trajectory to an
unstable periodic orbit can create a long line in the recurrence
plot. Our algorithm sytematically eliminates the possibility of
unstable periodic orbits, as one can easily see: A periodic time
series, of whatever large period length $T$, would result in
$U(x_{N+1})=0$ in Eq.(1) as soon as $N+1\ge T$, hence, the
periodic continuation is forbidden by our algorithm.

An AD system tries not only not to repeat the past trajectory but
to {\it escape} from it to other points of the phase space. It
follows that although for chaotic systems the attractor dimension
is finite and fixed for the AD motion the dimension of an observed
trajectory increases with the number of generated points. In this
sense the AD systems  are infinite dimensional  and  Unstable
Periodic Orbits cannot exist. Neigbouring trajectories  in AD
systems diverge  in all directions and  as result there  is  no
system attractor, i.e., the data are confined to a finite volume
in space only through the constraints (here: $x_i\in [-0.5:0.5]$).

\section{Time Series Analysis}
\par
The AD data share many properties of white noise, but there are
some which do not appear in any another dynamical behaviour.   Let
us first focus on features common with noisy data. It is easy to
check that our AD data converge to the uniform density in
the unit interval. The autocorrelation function, the mutual
information parameter as well as the power spectrum are for the AD
data of Eq.(1) the same as for the white noise, i.e., AD data
appear to be uncorrelated.

\begin{figure}
\begin{flushleft}
\includegraphics[scale=0.35,angle=-90]{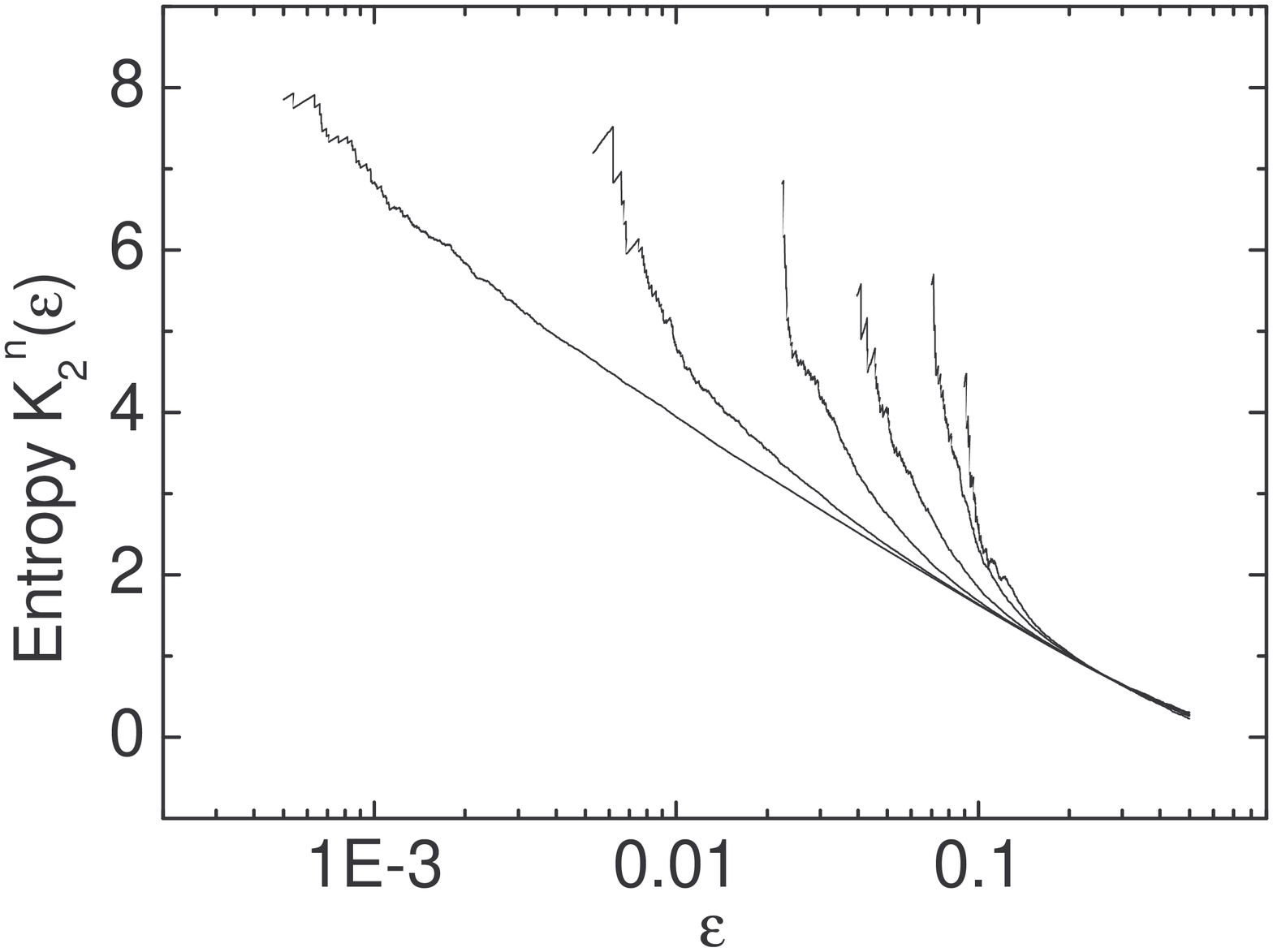}
\end{flushleft}
 \begin{flushleft}\vspace{-3.7cm}\hspace{1.2cm}\subfigure{
\includegraphics[scale=0.23]{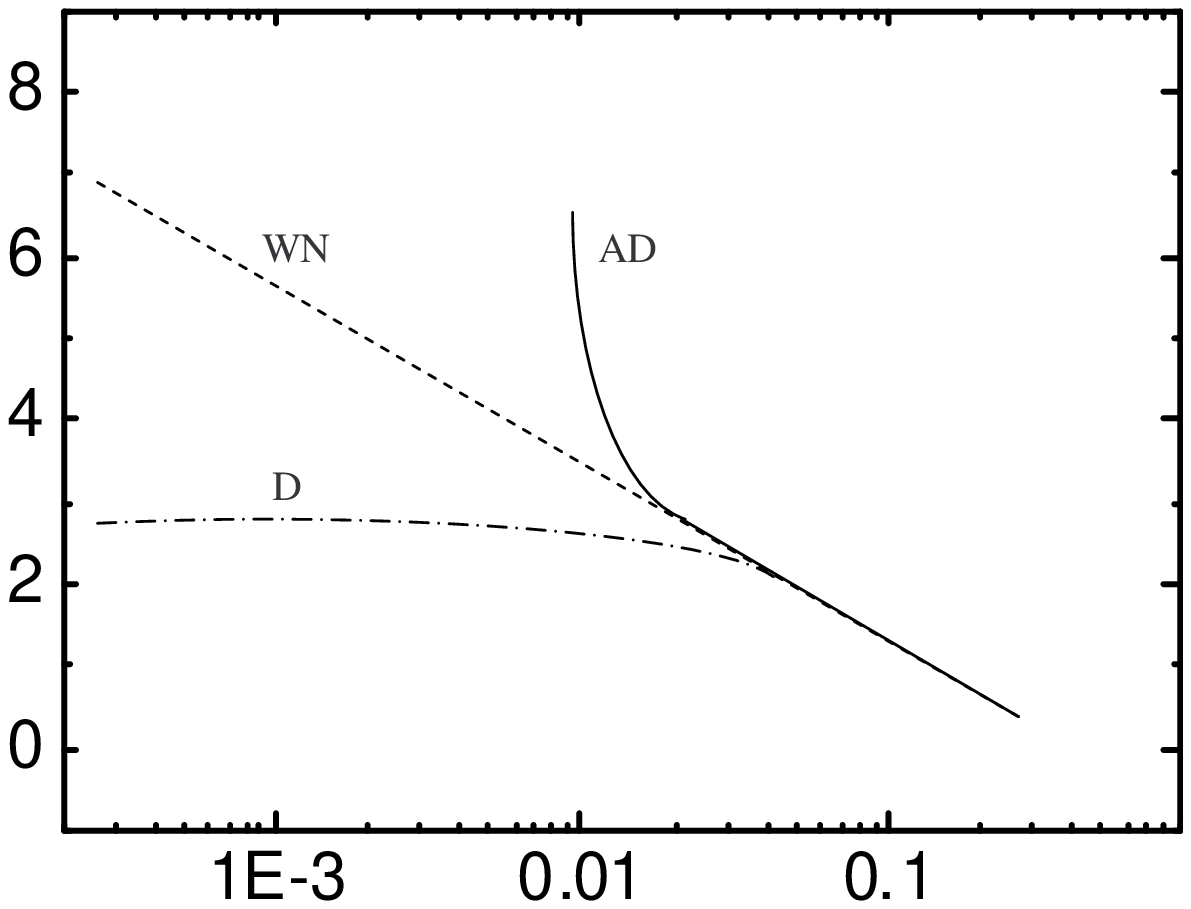}
}\end{flushleft}\caption{\label{fig.entropy} Coarse-grained block
correlation entropy calculated for AD systems with $d=2$, $m=4$.
Curves from the left to right correspond to embedding dimensions
increasing from $n=2$ to $n=7$ while the straight line corresponds
to entropy of white noise. The inset shows schematically the
behaviour of the coarse-grained correlation entropy for different
types of motion: (dashed dotted line) deterministic systems (D),
(dashed line) white noise (WN) and (solid line) AD.}
\end{figure}

Of course, by construction, there are subtle long range
correlations which have to show up in a suitable analysis.
The block correlation entropies $K_2^n$ offer one such analysis.
In terms of RP a line of length $1$ in  an embedding
dimension $n$  and a line of length $n$ in an embedding dimension $1$ are
equivalent \cite{Faure}. For this reason  one can alternatively use  terms
{\it embedding dimension} and {\it length of a line} (the embedding
dimension used in
RP is one  for all the calculations). Let us define the number of
lines in RP of length $n$ or longer as $DET_n(\eps)$ \cite{urb03}. Then
the coarse-grained block correlation entropy can be defined as
\cite{Proccacia} \be K_2^n(\eps)=\ln \frac{ DET_n(\eps)}{
DET_{n+1}(\eps)}\end{equation} and it corresponds to the slope in the plot $\ln
DET_n$ versus $n$. The differences between the noise and AD data can be
seen very clearly when we plot the coarse-grained block correlation
entropy $K_2^n$ versus the embedding dimension $n$ (see
Fig.~\ref{fig.entropy}). The entropy of AD data measured from numbers
$DET_n(\eps)$ of lines of
 length $n=3$ or longer is higher than the white noise entropy which in
Fig.~\ref{fig.entropy} is a straight line  with the slope
$-\ln(10)$. The minimal threshold for which the entropy can be
calculated  for $n=3$ corresponds to the maximal $\eps_{max}$ for
which there are no lines in RP of the length $n=3$ or longer. The
maximal entropies which can be obtained from a finite time series
of white noise and of the AD time series of the same length are
the same, but the maximum for AD data is always reached with
larger values of $\eps_{max}$ as compared to noisy data.

\par
\begin{figure}
\includegraphics[scale=0.35,angle=-90]{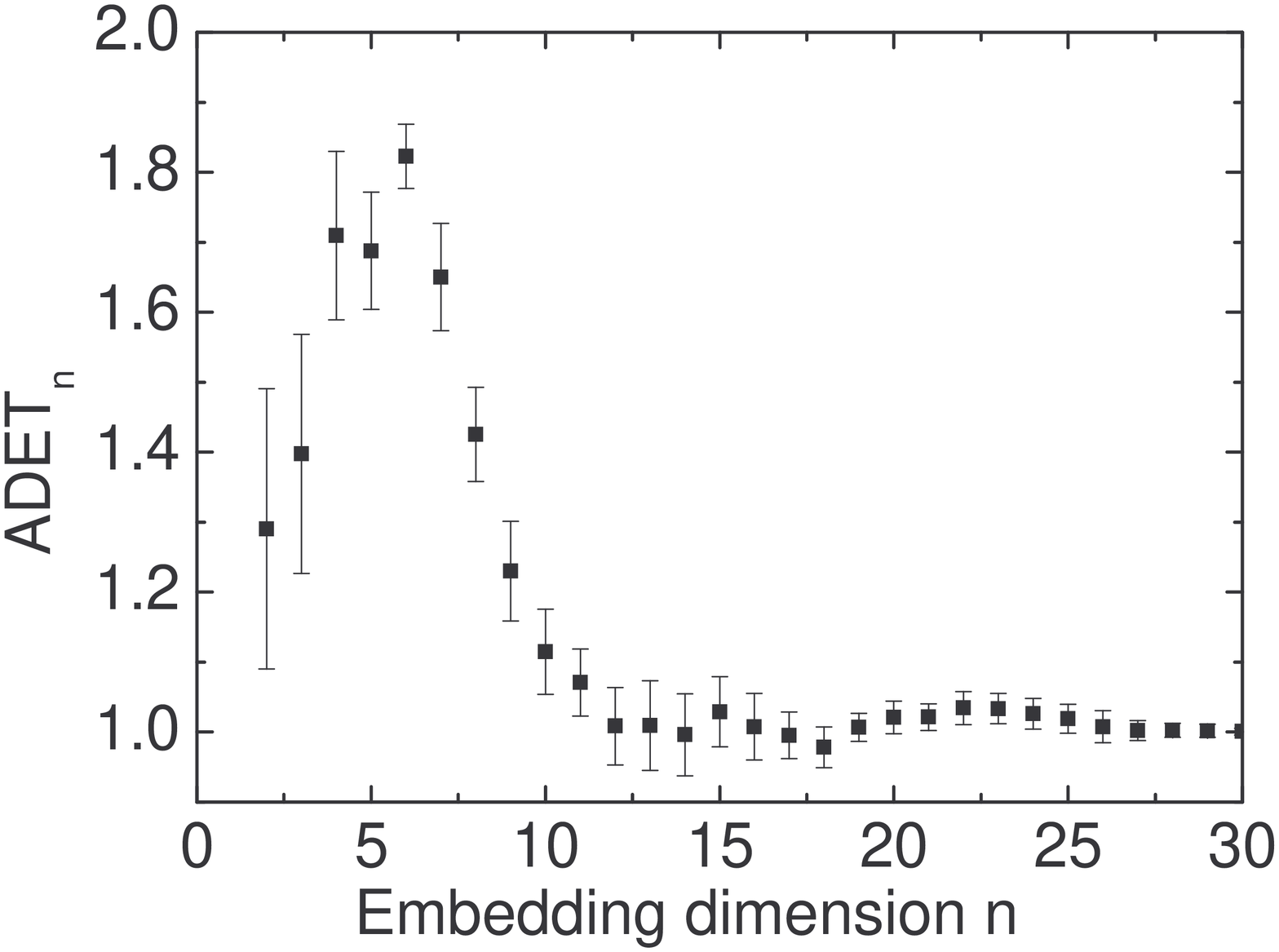}
\caption{\label{fig.ADET} The  plot of parameter $ADET_n$ versus
$n$ ($d=2$ and $m=4$).}
\end{figure}

 The same  feature of AD data  can  be analyzed from another point
of view. We have calculated the
minimal distance between nearest neighbors for the whole data set and
have divided it by the same quantity of shuffled data. Because for random
data the minimal distance between nearest neighbors can differ for
different shuffling an appropriate  averaging has been performed. In such a way
we create a parameter that we call $ADET_n$ which is a measure how
much data are anti-deterministic ($n$ is
here an embedding dimension). We have checked that for chaotic data
with noise and for large $n$
 this parameter converges  to the noise level
\be
NTS=\frac{\sigma_{noise}}{\sigma_{DATA}}\;,
\end{equation}
where
$\sigma_{noise}$ is
the standard deviation  of noise and $\sigma_{DATA}$ stands for the
standard deviation of data. Hence, for (noisy) deterministic data it is
usually much smaller than unity. Shuffling makes no difference for random
data and hence $ADET_n=1$. Fig.~\ref{fig.ADET} shows a plot of values $ADET_n$
versus $n$  for AD data with $d=2$ and $m=4$. One can see that for
$n$ between $4$ and $7$   we have $ADET_n \approx 1.75 $ what means
that the mean minimal distance
between nearest neighbors for AD data is about $1.7$ times larger than
for  noisy data, and very much larger than for conventional
deterministic data.

\par
\begin{figure}
\includegraphics[scale=0.35,angle=-90]{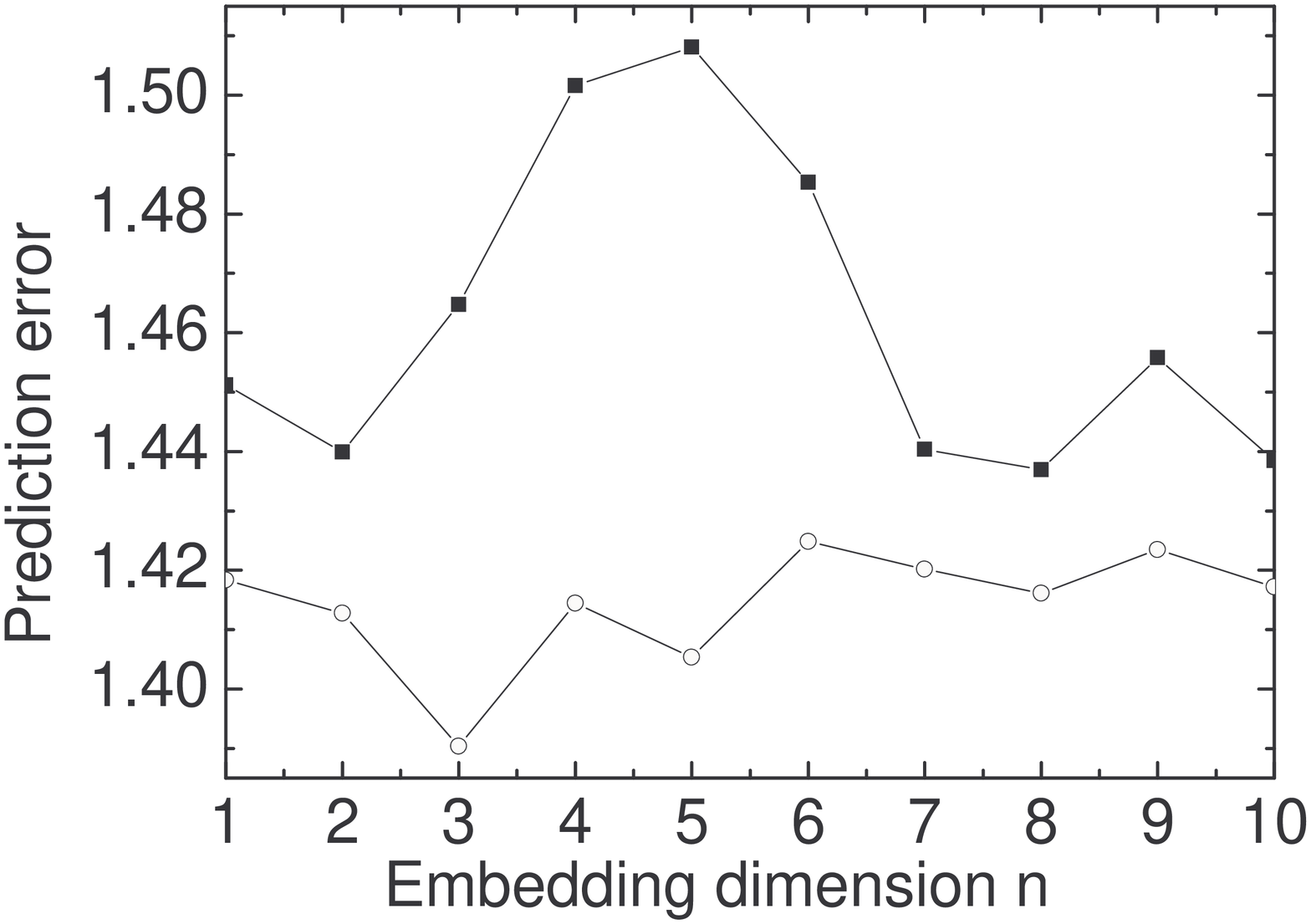}
\caption{\label{fig.pred} The prediction error divided by standard
deviation of data for AD data ($d=2$, $m=4$) and for randomly shuffled
data.}
\end{figure}

 To  demonstrate that AD data are {\it less predictable} than noise
we have calculated a
prediction error that one receives  from a zeroth order
prediction\cite{FarmerSidorowich}
using the nearest
neighbor. This simple forecasting method shows clearly that our
AD data are less predictable than noise, i.e.,  than randomly shuffled
data (see Fig.~\ref{fig.pred}). The average
prediction error for noisy data should
equal $\sqrt{2}$ because we have the root mean square of a quantity
which is the sum of two random numbers,
and the error of our AD data is larger for a range of
embedding dimensions.

Long range correlations in data may also express themselves in
anomalous diffusion properties when integrating the
data\cite{DFA}. This can be quantified by the Hurst exponent. The
Hurst exponent of data generated by Eq.(1) is found to be 1/2,
which is the value of white noise. However, using a different,
more complicated algorithm (see appendix~\ref{sec.recipe2}), we
were able to create data which are anti-persistent, i.e., for
which we found Hurst exponents ranging from $0.15$ to $0.5$,
depending on parameters in the algorithm. Observing histograms of
such data we have found that for finite time series lengths, the
data can be much more uniformly distributed than for standard
noise generators (used commonly in C compilers). Thanks to this
property one can generate very short time series with a very flat
histogram.

\section{Applications}
There may be different useful applications of AD data. One of them
are  noise generators. Common features of  AD data and noisy data
are in this case very desired. Some of the properties can be
modified changing the method and its parameters, e.g., to get a
desired  anti-persistence. The big advantage here might be given
by the property that histograms converge much faster to their
limiting shape than for uncorrelated random numbers, so that
smaller samples might yield sample means which are closer to the
true mean than for truly random samples. In Monte Carlo methods
for searching global minima, the feature of having large minimal
distances might enhance the efficiency due to the fact that it is
less likely to converge to the same local minimum with another
initial condition from the sample.

The AD systems can also be used in data encryption because {\it
noisy} behavior makes unwanted deciphering very difficult. As we
mentioned, there are many ways of how to modify the method
generating AD data which can produce entirely different time
series.  One can imagine that an encryption key can be parameters
of a modificated scheme.

We suspect that AD behaviour can be
observed in nature as well.  One evident possibility is to see it as
results of  {\it games}  when an intelligent
player tries to find a strategy that would be as much as possible
unpredictable to an intelligent opponent. This can occur in
predator-prey relations as well as in economical processes.
In physics, adding one
by one charged particles to a set of fixed charges with the
requirement of consuming minimal energy would result in a sequence
of positions of the added particles which is generated by a
functional similar to Eq.(1),
$U_x=\lb\sum_{i=1}^N\frac{1}{\abs{x_i-x_{N+1}}}\rb^{-1}$ , but
working in a, say, 2-dimensional space.

\section{Conclusions}
\par We present  a new type of system behavior that is neither
periodic, nor quasiperiodic,  chaotic or stochastic in a
conventional sense. This kind of motion violates the main feature
of common deterministic systems, i.e., occurrence of  lines
parallel to the main diagonal in  recurrence plots. We call this
kind of motion anti-deterministic. AD systems have many common
features with white noise. Nonlinear time series analysis reveals
that anti-determinism is related to a divergent behaviour of block
correlation entropies computed on finite data set, which is
consistent with the observation that they are less predictable
than white noise.

\begin{acknowledgments}
JAH is thankful to the Complex Systems Network of Excellence {\it
  EXYSTENCE}  for the financial support that
made possible his visit at MPI-PKS Dresden.
\end{acknowledgments}
\par
\appendix
\section{\label{sec.recipe2}{Another algorithm for AD data generation}}
\par  For each
$i=1,2,\ldots,N-1$ we calculate in dimension $d$ the square of the
Euclidean distance $dis_d\lb i,N\rb$ between the point $x_i$ and
$x_N$: \be dis_d(i,N)=\sum\limits_{k=0}^{d-1} \lb
x_{i-k}-x_{N-k}\rb^2\;. \ee Then we determine the following
parameter $P_{k}$ created from the above distances \be
P_{k+1}=\max\limits_{d=\kappa,\kappa+1,\ldots,k}\set{\frac{(d)^q}{dis_d\lb
k,N\rb}}.\label{eq.utilP2}\ee Here $\kappa$ and $q$ are the
parameters of the method, and the maximum over $d$ is usually
obtained for $d$ not much bigger than $\kappa$. Now we have to
discretize the domain of the set $\set{x_i}$ on $S$ intervals
$\set{z_k}$ for $k=1,2,\ldots,S$, so we  have
$z_k=(min_x+(k-1)\Delta x,min_x+k\Delta x]$ where $\Delta
x=\frac{max_x-min_x}{S}$. Here $max_x$ is the maximal value from
the set $\set{x_i}$ and $min_x$ is the minimal value respectively.
Next for each interval $z_k$ we calculate the utility function
$U_k$ as follows \be U_k=\sum\limits_{i=\kappa+1}^{N} P_{i}\cdot
g_b\lb\abs{N\frac{x_i-min_x}{max_x-min_x}-k}\rb.\label{eq.utilU}
\ee In our case we use the hyperbolic function \be g_b\lb
y\rb=\frac{1}{y+b}\label{eq.g},\ee for $b>0$. $b$ appears as the
parameter of the method. The last step is to look for the index at
the minimal utility function $m=\set{k:\min\limits_k \set{U_k}}$
and create variable belonging to the interval $z_m$. We determine
the exact solution as follows \be x_{N+1}=
\begin{cases}
    m+1-\frac{U_{m+1}-U_m}{U_{m-1}-U_m} & U_{m+1}<U_{m-1}, \\
    m-1+\frac{U_{m-1}-U_m}{U_{m+1}-U_m} & \text{otherwise}.
  \end{cases}\ee
After we found the value of $x_{N+1}$  we repeat the whole
procedure for the larger set $\set{x_i}$ where $i=1,2,\ldots,N+1$
and so on. Main differences of this algorithm for AD data
generation to the previous one is that we use a range of
dimensions $d$ in Eq.~(\ref{eq.utilP2}) and a global influence of
every point $i$ to the utility function~(\ref{eq.utilU}).
%\newpage
%\bibliography{apssamp}

\end{document}